\documentclass[onecolumn,showpacs,preprintnumbers,amsmath,amssymb]{revtex4-2}

\usepackage{graphicx, amsmath, amssymb, amsfonts, mathtools, mathrsfs, color}
\usepackage{natbib, comment, todonotes}
\usepackage[margin = 1.5cm]{geometry} 
\usepackage{xcolor}
\usepackage{soul}
\usepackage{dcolumn}
\usepackage{xcolor}
\usepackage[capitalise]{cleveref}
\usepackage[normalem]{ulem}
\usepackage{bm}

\newcommand{\td}[2]{\frac{d #1 }{d #2}}

\newcommand{\pd}[2]{ \frac{ \partial #1}{ \partial #2 } }

\newcommand{\pdi}[2]{ \partial_{#2} #1 }

\newcommand{\bvec}[1]{\ensuremath{\boldsymbol{#1}}}
\newcommand{\grad}{\nabla}

\newcommand{\mean}[1]{ \langle #1 \rangle}

\newcommand{\Pra}{\text{Pr}}
\newcommand{\Ra}{\text{Ra}}
\newcommand{\Rey}{\text{Re}}
\newcommand{\Nu}{\text{Nu}}

\newcommand{\si}{\text{Supplemental Material}}

\newcommand{\uu}{\bvec{u}}

\newcommand{\Lap}{\grad^2}

\newcommand{\Area}{\text{A}_0}

\newcommand{\len}{h}
\newcommand{\er}{\bvec{e_r}}
\renewcommand{\eth}{\bvec{e_\theta}}

\newcommand{\xc}{X}
\newcommand{\yc}{Y}
\newcommand{\dy}{\Delta y}

\newcommand{\phim}{\phi_{\text{max}}}
\newcommand{\Ebot}{E_{\text{bot}}}
\newcommand{\Lmax}{L_{\text{max}}}

\begin{document}


\title{A convective fluid pendulum revealing states of order and chaos}
\author{Jinzi Mac Huang$^{1,2}$}
\email{machuang@nyu.edu}
\thanks{The authors contributed equally.}
\author{Nicholas J. Moore$^3$}%
\email{nickmoore83@gmail.com}
\thanks{The authors contributed equally.}
\affiliation{1. NYU-ECNU Institute of Physics and Institute of Mathematical Sciences, New York University Shanghai, Shanghai, 200122, China \\
2. Applied Math Lab, Courant Institute, New York University, New York, NY 10012, USA \\
3. Department of Mathematics, Colgate University, Hamilton, NY 13346, USA
}%
\begin{abstract}
We examine thermal convection in a two-dimensional annulus using fully resolved direct numerical simulation (DNS) in conjunction with a low-dimensional model deriving from Galerkin truncation of the governing Navier-Stokes Boussinesq (NSB) equations. The DNS is based on fast and accurate pseudo-spectral discretization of the full NSB system with implicit-explicit time stepping. Inspired by the numerical results, we propose a reduced model that is based on a Fourier-Laurent truncation of the NSB system and can generalize to any degree of accuracy. We demonstrate that the lowest-order model capable of satisfying all boundary conditions on the annulus successfully captures reversals of the large-scale circulation (LSC) in certain regimes. Based on both the DNS and stability analysis of the reduced model, we identify a sequence of transitions between (i) a motionless conductive state, (ii) a state of steady circulation, (iii) non-periodic dynamics and chaotic reversals of the LSC, (iv) a high Rayleigh-number state in which LSC reversals are periodic despite turbulent fluctuations at the small scale. The reduced model reveals a link to a damped pendulum system with a particular form of external forcing. The oscillatory pendulum motion provides an accurate prediction for the LSC reversal frequency in the high Rayleigh-number regime.
\end{abstract}

\date{\today}
\maketitle

\section{Introduction}

The nonlinear coupling between temperature and flow fields in thermal convection leads to a range of highly nontrivial dynamics \cite{Busse1978, camassa1991transport, Childress2009, Ahlers2009, McCurdy2019, McCurdy2022}. For example, the collective motion of a turbulent flow may form a large-scale circulation (LSC) that can drive atmospheric and oceanic patterns \cite{Salmon1998, Zhong2009}. The direction of the LSC is known to reverse \cite{Araujo2005, Brown2007}, which can lead to observable effects such as a sudden changes in wind direction \cite{Doorn2000}. Reversals of the LSC in mantle convection may even play a role in reversals of Earth's magnetic dipole \cite{Glatzmaier1999}, and in solar convection the magnetic switchbacks of the Sun \cite{Wit2020}. In other contexts, the mixing of moving fluids can substantially enhance heat transfer rates \cite{Ahlers2009, Chong2017, Belmonte1994, Grossmann2000, Niemela2000, Caves2019}, making fluids perfect coolants. It is even more interesting when thermal convection couples to a moving interface, where processes like melting \cite{Moore2017, favier2019rayleigh, Weady2022}, erosion \cite{Meakin2010, Ristroph2012, Moore2013, Quaife2018, Chiu2020, Moore2023PhysD}, dissolution \cite{Maruyama2000, Huang2015, Huang2020a, Huang2022}, and sublimation \cite{Bergeron2006, Claudin2015} are accelerated by convection as solid morphology evolves. On an extremely large scale, convection in the Earth's mantle drives plate tectonics \cite{Whitehead1972, Zhong2005, Whitehead2015, mac2018stochastic}. 

\begin{figure}[htb]
 \includegraphics[width=6in]{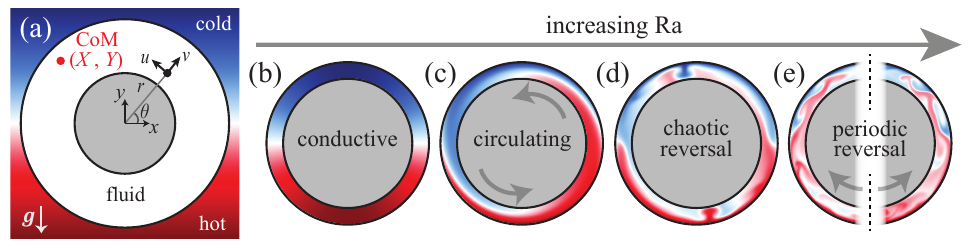}
  \caption{Thermal convection in an annulus. (a) Schematic showing the fluid domain [$r\in(r_0,1/2)$] as an annulus embedded in a solid with constant vertical temperature gradient. (b) At low $\Ra{}$, the fluid stays motionless and heat simply conducts through the fluid. (c) Increasing $\Ra{}$ beyond a critical value leads to circulating fluid motion. (d) Further increasing $\Ra{}$ leads to chaotic circulations that switch direction randomly. (e) At very high $\Ra{}$, the reversal motion becomes periodic while the small-scale flow is turbulent. Movies of (b)-(e) are included in the \si{}. Rayleigh-number values are $\Ra{} = 1.4\times10^5$ (b), $3.1\times10^6$ (c), $2.5\times10^7$ (d), and $1.1\times10^9$ (e). In all cases, $\Pra = 4$ and $r_0 = 0.4$.
  }
\label{fig1}
\end{figure}

In this manuscript and companion letter \cite{PRL2023}, we consider a canonical example of two-dimensional (2D) thermal convection shown in \cref{fig1}(a) \cite{Tritton1988}, in which an annular domain of fluid is heated from below. Depending on the strength of the thermal forcing, the fluid may remain motionless in a conductive state [\cref{fig1}(b)], circulate steadily in either the clockwise (CW) or counter-clockwise (CCW) direction [\cref{fig1}(c)], or exhibit reversals of the LSC, be they chaotic [\cref{fig1}(d)] or periodic [\cref{fig1}(e)]. 
The circulatory nature of thermal convection, which appears generically across many settings, is especially reinforced by the annular shape of the domain considered here. The feedback with geometry tends to create a single, dominant convection cell, while minimizing other effects, such as corner rolls or detached plumes \cite{Sugiyama2010, Ni2015, Araujo2005}. This large convection cell represents large-scale coherent motion observed at high Rayleigh number in other geometries and in other applications \cite{niemela2003rayleigh, Araujo2005, Brown2007, Ahlers2009, Wang2018}.

Previous studies have qualitatively linked the rich dynamics in \cref{fig1} to the famous Lorenz system \cite{Creveling1975, Kaplan1979, Singer1991}, a dynamical system describing thermal convection in a planar domain. Diverse states with order and chaos have been observed in both models \cite{Creveling1975, Kaplan1979, Gorman1984} and experiments \cite{futterer2007thermal, Gorman1986}. Phenomenological models developed for other geometries lend additional insight \cite{Araujo2005, Brown2007, Ni2015}. However, a precise model for annular convection that derives systematically from the governing equations and that quantitatively predicts the sequence of transitions is still missing. In this work, we revisit this classic configuration of thermal convection and analytically derive a low-dimensional model from the flow and heat equations. By studying the time evolution of the fluid center of mass (CoM) and the angular momentum $L$, our work links the thermal convection in \cref{fig1} to a mechanical pendulum problem for CoM and provides analytic predictions of the onset of fluid motion and chaos.

In the following, we introduce the governing equations (Sec. \ref{sec-gov}), and then discuss a  scheme to perform direct numerical simulations of thermal convection (Sec. \ref{sec-num}). We next link the observed nonlinear behaviors to a low-dimensional system of ordinary differential equations (ODEs) (Sec. \ref{sec-lowdim}). The derivation of this dynamical system requires minimal assumptions, and it recovers the numerical observations over a large range of parameters. Through analyzing the critical states and stability of the ODE system, the transition between different dynamical states can be analyzed in detail (Sec. \ref{sec-bir}). We finally show how the high Rayleigh-number convection can be linked to a mechanical pendulum (Sec. \ref{sec-pendulum}), and close with some concluding remarks (Sec. \ref{sec-discussion}).

\section{Governing equations}
\label{sec-gov}

Consider an annular fluid domain of inner radius $R_0$, outer radius $R_1$, and height $\len = 2 R_1$ as depicted in Fig.~\ref{fig1}(a). The fluid is heated from below, with temperature difference $\Delta T$ between the bottom and top of the annulus. Material properties include the kinematic viscosity $\nu$, the thermal diffusivity $\kappa$, the thermal expansion coefficient $\beta_T$, and the average fluid density $\rho_0$. The coupled fluid flow and heat transport inside the annulus are governed by the incompressible Navier-Stokes equations with Bousinesq approximation (NSB):
\begin{align}
\label{NS}
& \pd{\uu}{t} + \uu \cdot \grad \uu = -\grad p +
\Pra \Lap \uu + \Pra \, \Ra \, T \bvec{e_y}, \\
\label{trans}
& \pd{T}{t} + \uu \cdot \grad T = \Lap T, \\
\label{incomp}
& \grad \cdot \uu = 0.
\end{align}
The above equations are in dimensionless form, with space scaled on $\len$, time on $\len^2 / \kappa$ (the thermal diffusive timescale), velocity on $\kappa / \len$, and density variations on $\Delta \rho = \rho_0 \beta_T \Delta T$. Variables include the dimensionless velocity $\uu$, pressure $p$, and temperature $T$ fields. \Cref{NS,trans,incomp} represent conservation of momentum, thermal transport, and  incompressibility respectively. Dimensionless parameters include the Rayleigh number $\Ra$ and the Prandtl number $\Pra$: 
\begin{align}
\label{RaPra}
\Ra = \frac{\beta_T \Delta T \len^3 g}{\nu \kappa} \, , \quad
\Pra = \frac{\nu}{\kappa}  .
\end{align}
The Rayleigh number measures the relative strength of thermal forcing, while the Prandtl number measures the ratio of momentum to thermal diffusivity.

The imposed temperature on the outer boundary of the annulus decreases linearly with height, while the inner boundary remains adiabatic. Meanwhile, the velocity field, expressed as $\uu = u \eth + v \er$ in polar coordinates, satisfies no-slip conditions on both boundaries. The boundary conditions are thus:
\begin{align}
\label{noslip}
&u = v = 0 \hspace{25pt} \text{at } r=r_0 \text{ and } r=1/2, \\
\label{Tinner}
&\pd{T}{r} = 0 \hspace{35pt} \text{at } r=r_0, \\
\label{Touter}
&T = \frac{1-\sin \theta}{2} \hspace{10pt} \text{at } r=1/2 .
\end{align}
Due to our nondimensionalization, $r = 1/2$ represents the outer boundary and $r = r_0 = R_0 / 2R_1$ the inner boundary. In \cref{trans,NS,incomp,RaPra,noslip,Tinner,Touter}, the three dimensionless numbers $\Ra, \Pra, r_0$ serve as the control parameters.

We first note that system Eqs.~\eqref{NS}--\eqref{Touter} supports a conductive state in which the temperature decreases with height and no fluid motion occurs. By setting $\uu = 0$ in \cref{trans} and satisfying boundary conditions \cref{Tinner,Touter}, the conductive-state temperature distribution can be determined in exact form as 
\begin{equation}
\label{conductive}
T_{\text{cond}} = \frac{1}{2} - \frac{r_0}{1+4r_0^2} \left(\frac{r}{r_0}+\frac{r_0}{r}\right)\sin{\theta}.
\end{equation}
In this state, relatively cool, heavy fluids sits atop warm, light fluid, thus raising the center of mass (CoM). Notice that, as a consequence of the inner adiabatic condition, the temperature within the fluid does not simply vary {\em linearly} with height. The dimensionless coordinates of the CoM are given by
\begin{align}
\label{COM}
\xc = - \frac{1}{\Area} \int_{\Omega} x \, T \, dA \, , \quad
\yc = - \frac{1}{\Area} \int_{\Omega} y \,  T \, dA .
\end{align}
where $\Area = \pi(1-4r_0^2)/4$ is the area of the annulus and $dA = r\,dr d\theta$ is the area element. We note that the dimensional CoM can be obtained by multiplying by $\len \beta_T \Delta T$. Inserting \eqref{conductive} into \eqref{COM} and integrating, gives the height of the conductive-state CoM,
\begin{equation}
\label{y0_cond}
y_{0} = \frac{ 1+12 r_0^2 }{16 (1+4r_0^2)} .
\end{equation}
From this formula, it is clear that $y_0 > 0$ for any value of $r_0$, thus confirming that the conductive-state CoM lies above the center of the annulus. 

At sufficiently high Rayleigh number, the top-heavy conductive state gives way to thermal convection. The circulatory nature of thermal convection, which appears generically across many settings, is especially reinforced here by the annular shape of the domain. This feedback with geometry tends to create a dominant convection cell that fits the annulus, although fine-grained, complex dynamics may appear in combination. To characterize the leading-order dynamics, we introduce the {\em average angular momentum} $L$ of the fluid
\begin{equation}
\label{Ldef}
L = \frac{1}{\Area}\int_0^{2\pi} \int_{r_0}^{1/2} r^2 u \, dr d\theta .
\end{equation}
Here, $L>0$ corresponds to rotation in the counter-clockwise (CCW) direction.

\section{Direct numerical simulations}
\label{sec-num}

In this section, we discuss direct numerical simulation (DNS) of the NSB system given by \cref{NS,trans,incomp,RaPra,noslip,Tinner,Touter}. We first introduce the numerical methods and then discuss results of the simulations.

\subsection{Numerical methods}

Our simulation method is based on the 2D streamfunction-vorticity form of \cref{NS,trans,incomp}:
\begin{align}
\label{omegaeq}
\pd{\omega}{t} + \uu \cdot \grad \omega &= \Pra \Lap \omega   + \Pra\, \Ra\, \left(\frac{\partial T}{\partial r}\cos\theta-\frac{1}{r}\frac{\partial T}{\partial \theta}\sin\theta\right),\\
\label{Teq}
\pd{T}{t} + \uu \cdot \grad T &= \Lap T,\\
\label{psieq}
 -\Lap \psi = \omega,\ \  \uu &= \nabla_\perp \psi.
\end{align}
Rather than solving for $\uu$ and $p$, our method solves for the vorticity $\omega = r^{-1} \left[\partial_r(r u)-\partial_\theta v\right]$ 
and stream function $\psi$. Velocity can then be recovered as $\uu  = \nabla_\perp \psi = r^{-1}\psi_\theta\er - \psi_r\eth$, so $u= - \psi_r$ and $v = r^{-1}\psi_\theta$. 

We first discretize time with the second-order Adam-Bashforth Backward Differentiation method (ABBD2). At time step $t = n\Delta t$, \cref{omegaeq,Teq,psieq} become
\begin{align}
\label{omegadisc}
\Lap \omega^{(n)} -\sigma_1\omega^{(n)}  &= f^{(n)},\\
\label{Tdisc}
\Lap T^{(n)} -\sigma_2 T^{(n)}  &= g^{(n)},\\
\label{psidisc}
 -\Lap \psi^{(n)} &= \omega^{(n)},
\end{align}
where
\begin{align}
\Lap = \frac{\partial^2}{\partial r^2} + \frac{1}{r}\pd{}{r} + \frac{1}{r^2} \frac{\partial^2}{\partial\theta^2}, \quad \sigma_1 = &\frac{3}{2\,\Pra\,\Delta t},\quad \sigma_2 = \frac{3}{2\Delta t}, \\[8pt]
f^{(n)} = \Pra^{-1}\left[ 2  (\uu \cdot \grad \omega)^{(n-1)} - (\uu \cdot \grad \omega)^{(n-2)}\right] &- (2\, \Pra\, \Delta t)^{-1} \left(4\omega^{(n-1)}-\omega^{(n-2)}\right) \\
&-\Ra\, \left(T_r\cos\theta-r^{-1}T_\theta\sin\theta\right)^{(n)},\notag \\[8pt]
g^{(n)} = \left[ 2  (\uu \cdot \grad T)^{(n-1)} - (\uu \cdot \grad T)^{(n-2)}\right] &- (2\Delta t)^{-1} \left(4T^{(n-1)}-T^{(n-2)}\right) .
\end{align}

ABBD2 is an implicit-explicit (IMEX) method for solving the stiff advection-diffusion equations, where the diffusion is handled by the backward differentiation method and the advection terms are handled by the Adam-Bashforth method. Furthermore, explicit and nonlinear terms in $f^{(n)}$ and $g^{(n)}$ are computed pseudo-spectrally, with an anti-aliasing filter detailed in \cite{Hou2007}. Through properly arranging the IMEX operator splitting, the overall accuracy of this method is second order in time. This well-tested method has been implemented in various convection problems \cite{peyret2002spectral,mac2021stable,Huang2022a}, yielding accurate solutions for a wide range of parameters.

\Cref{omegadisc,Tdisc,psidisc} are Helmholtz and Poisson equations that can be solved by standard numerical methods. Considering that $r\in(r_0,1/2)$ and $\theta$ is periodic, we discretize $r$ variable with a Chebyshev series and $\theta$ variable with a truncated Fourier expansion. 
For more details of this Fourier-Chebyshev method, see \cite{peyret2002spectral,Huang2022a}. With given initial and boundary data, \eqref{Tdisc} can be solved first to obtain $T^{(n)}$, which is inserted in $f^{(n)}$ so \eqref{omegadisc} can be solved next. Finally, \eqref{psidisc} is solved with the known $\omega^{(n)}$.

Typically, we use 1024 Fourier modes and 128 Chebyshev nodes in our simulation, which yields resolved and accurate solutions. We further set $\Delta t = 5\times 10^{-4} \,\Ra^{-1/2}$ to maintain time-stepping accuracy and stability, considering that the characteristic flow speed scales as $|\uu|\sim \sqrt{\Ra}$ (to be shown later).

The main difficulty of solving \cref{omegadisc,Tdisc,psidisc} is from the boundary conditions,
\begin{align}
\psi=\psi_r &= 0,\quad T_r = 0 \quad&\mbox{at } r = r_0,\\
\psi = -Q(t),\quad \psi_r &= 0,\quad T = \frac{1-\sin \theta}{2} \quad  &\mbox{at } r = \frac{1}{2}.
\end{align}
In these boundary conditions, we have both Neumann and Dirichlet data on $\psi$, but no boundary data on $\omega$. This situation can be handled by the influence matrix method \cite{peyret2002spectral}, which is a method to numerically map the Neumann data of $\psi$ to the Dirichlet data of $\omega$. 

Due to the nonzero flow circulation, we also have to determine the flux $Q(t) = (2\pi)^{-1}\int_0^{2\pi}\int_{r_0}^{1/2} u(r,\theta, t)\, drd\theta$ in the Dirichlet data of $\psi$. Denoting the $\theta$ average of $f$ as $\widehat{f} = (2\pi)^{-1}\int_0^{2\pi}f\, d\theta$ and averaging the $\eth$ component of \cref{NS}, we have
\begin{align}
    \label{uhateqn}
    \frac{\partial \widehat{u}}{\partial t} +   \widehat{(u_r v)} + \frac{\widehat{(u v)}}{r}  &= \Pra\, \Ra\, \widehat{(T\cos{\theta})} + \Pra\, \left(\frac{\partial^2 \widehat{u}}{\partial r^2} + \frac{1}{r} \frac{\partial \widehat{u}}{\partial r} - \frac{\widehat{u}}{r^2}  \right),\\
    \widehat{u}(r_0,t) &= \widehat{u}(0.5, t) = 0.
\end{align}

At time $t = n\Delta t$, values of $u^{(n-1)}$, $v^{(n-1)}$ are known and $T^{(n)}$ can be solved by \cref{Tdisc} first, therefore we can solve $\widehat{u}(r,t)$ pseudo-spectrally with Chebyshev method. Finally, $Q(t) = \int_{r_0}^{1/2} \widehat{u}(r,t) \, dr$.

\subsection{Numerical results}

In this section, we briefly present some results of the DNS and show how tuning parameters like $\Ra$ can lead to diverse dynamical states. In all simulations, we set the inner radius to be $r_0 = 0.4$ so that dynamics are confined to a relatively narrow annulus.  At $\Pra = 4$, numerically solving \cref{NS,trans,incomp,RaPra,noslip,Tinner,Touter} yields fluid motions shown in \cref{fig1}, with corresponding movies included in the \si{}. \Cref{fig1}(b) shows the low-$\Ra$ case, in which buoyancy is too weak to overcome viscous forces. In this conductive state, the fluid is motionless, and the only mechanism for thermal transport is conduction. As $\Ra{}$ increases, the destabilizing buoyancy becomes strong enough to drive a circulating flow shown in \cref{fig1}(c), where the fluid circulates unidirectionally in either the CW or CCW direction. At even higher $\Ra$, \cref{fig1}(d) shows that the flow is no longer unidirectional, but reverses between CW and CCW in a chaotic manner. Counterintuitively, this reversal becomes regular as $\Ra$ gets even higher, where the flow is turbulent but the bulk motion reverses periodically as shown in \cref{fig1}(e).

\begin{figure}[htb]
 \includegraphics[width=6.5in]{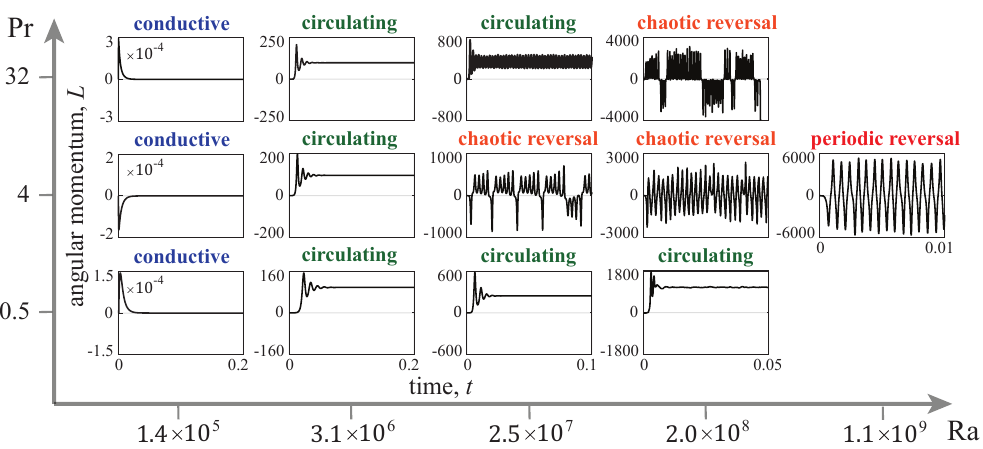}
  \caption{Angular momentum $L$ at various $\Ra{}$ and $\Pra$ reflecting the strength and direction of circulation. Four states of motion can be identified: conductive, circulating, chaotic reversal, and periodic reversal states. In all simulations, $r_0 = 0.4$. }
\label{fig2}
\end{figure}

To better capture the circulatory nature of flow in the thin channel, \cref{fig2} shows the time series of the angular momentum $L(t)$ as defined in \cref{Ldef}. At each $\Pra$, the general trend of conductive, circulating, and reversal flow patterns emerges as we increase $\Ra$, where the critical $\Ra$ separating these states differs depending on $\Pra$. At $\Ra{} = 1.4\times10^5$ (first column of \cref{fig2}), the motionless conductive state is a stable equilibrium for all the DNS presented, as initially added perturbation in $L$ decays rapidly. In this case, a steady state exists for \cref{NS,trans,incomp,RaPra,noslip,Tinner,Touter} which has no flow ($L=0$) and a conductive temperature profile given by \cref{conductive}.

The second column of \cref{fig2} shows the case $\Ra{} = 3.1\times10^6$, in which the flow reaches a steady circulating state with nonzero angular momentum. Although all the circulating states in \cref{fig2} have a CCW direction, we note that CW circulating flow is also possible, depending on the initial distribution of temperature and flow.  

So far, the flow and temperature profiles are  steady as $t\to\infty$, however this steadiness is lost as we increase $\Ra{}$ to $2.5\times10^7$ for $\Pra = 4$. This particular case shows that a steady circulating flow can also destablize, leading to the state of chaotic reversal, where the circulation is no longer unidirectional and the flow switches between CW and CCW directions. The dynamics of $L(t)$ are chaotic, as small perturbations to the initial condition lead to very different trajectories. We will later show that the Lyapunov exponent in this case is large.  

Although one might expect the state of chaos to persist, or even intensify, as the Rayleigh number increases, surprisingly order returns at sufficiently high $\Ra$.
At the Rayleigh number of $1.1\times10^9$ (last column of \cref{fig2}), the trajectory of $L(t)$ oscillates periodically, even though the flow structure is far from trivial as shown in the supplemental movies. How does this oscillatory state appear even though the flow is turbulent? What determines the frequency of the reversals? We address these questions in later sections through a simplified ODE model that links this periodic flow reversal to the oscillation of a mechanical pendulum.

The reversal states also exist for simulations with different $\Pra$ shown in \cref{fig2}, with an exception of those with $\Pra{}=0.5$ where the dynamics do not transition to chaos. Through investigation of the ODE model in later sections, we identify a critical Prandtl number $\Pra^*$, below which the state of steady circulation remains stable for arbitrarily large $\Ra$.

\begin{figure}[htb]
 \includegraphics[width=6.5in]{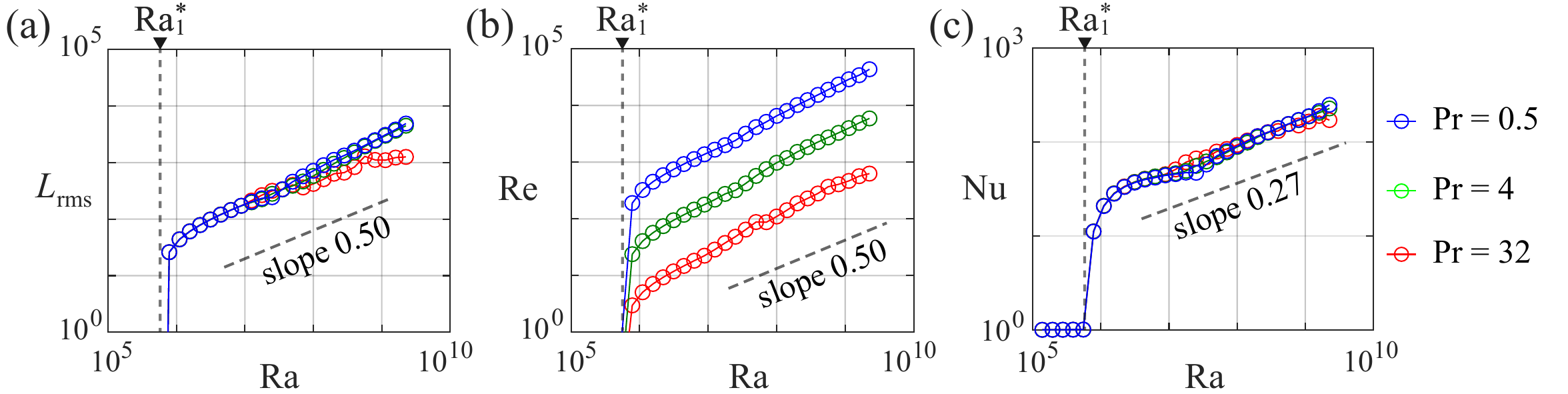}
  \caption{Time-averaged bulk quantities of the flow and temperature fields. (a) RMS value of angular momentum $L_\text{rms}$ represents the circulation strength. (b) Reynolds number $\Rey$ indicates the scale of flow speed. (c) Nusselt number $\Nu$ measures the heat passing through the fluid domain. In all simulations, $r_0 = 0.4$, and a common critical Rayleigh number can be identified as $\Ra_1^* = 7.25\times10^5$.}
\label{fig3}
\end{figure}

In \cref{fig2}, the scale of $L$ apparently depends on $\Ra$ and $\Pra$. As a measure of this scale, we define the root mean square of $L$ as 
\begin{equation}
    L_{\text{rms}} = \sqrt{\langle L^2 \rangle},
\end{equation}
where $\langle \cdot \rangle$ is the time average operation. \Cref{fig3}(a) shows that $L_{\text{rms}}$ remains zero for low $\Ra$ until a critical value $\Ra^*_1 = 7.25\times10^5$ (this value will be identified later), at which point $L_{\text{rms}}$ grows positive with increasing $\Ra$. That is, higher $\Ra$ results in stronger circulation. In the high $\Ra$ limit, $L_{\text{rms}}\propto\sqrt{\Ra}$ and has a weak dependence on $\Pra$. 


The flow velocity on the other hand, can be represented by the Reynolds number,
\begin{equation}
\label{Rey}
    \Rey =  \Pra^{-1} \langle \max |\uu|\rangle,
\end{equation}
where we use the maximum flow speed $\max |\uu|$ to represent the velocity scale and $\Pra{}$ to represent the scale of kinematic viscosity. We note that the definition \cref{Rey} is a consequence of our non-dimensionalization procedure, where we have rescaled the length by $h$ and speed by $\kappa/h$, so $\Rey = h \langle|U|\rangle/\nu = (\kappa/\nu) \max |\uu|$, where $U$ is the maximum dimensional flow speed.

Shown in \cref{fig3}(b), $\Rey$ also becomes nonzero as $\Ra>\Ra_1^*$, indicating the onset of fluid motion. At high $\Ra$, $\Rey$ also has a power-law dependence with $\Ra$ that has an exponent near $0.5$, agreeing with the value obtained from Rayleigh-B\'enard convection \cite{Ahlers2009, Huang2022a}. Interestingly, the scale of flow speed $\max|\uu|$ at a constant $\Ra{}$ is not strongly influenced by $\Pr$, as $\Rey$ at a fixed $\Ra$ is inversely proportional to $\Pra{}$ in \cref{fig3}(b).

After analyzing the flow structures, we now turn our attention to the heat transfer. To measure the amount of heat passing through the fluid, we define the dimensionless Nusselt number,
\begin{equation}
    \Nu = \frac{\langle q\rangle}{\langle q_{\text{cond}}\rangle} = \frac{\langle \int_0^\pi (\partial_r T)|_{r=1/2}\, d\theta\rangle}{\langle\int_0^\pi (\partial_r T_{\text{cond}})|_{r=1/2}\, d\theta\rangle}.
\end{equation}
Above, $q$ is the total heat flux measured in the DNS, while $q_{\text{cond}}$ is the heat flux associated with the conductive temperature field in \cref{conductive}. Naturally, $\Nu{} = 1$ for solids and motionless fluids, while thermal convection gives $\Nu{} > 1$, meaning the moving fluid is able to carry more heat convectively. Indeed, we observe this transition in \cref{fig3}(c), where a sudden increase of $\Nu{}$ can be spotted at $\Ra_1^*$. In the limit of high $\Ra$, a power law scaling $\Nu\sim\Ra^{0.27}$ emerges, similar to the scaling observed in the Rayleigh-B\'enard convection (i.e.~planar boundaries) \cite{Niemela2000, Ahlers2009, Huang2022a}.

With the simple geometry of an annulus, it becomes possible to analyze the flow and temperature dynamics of thermal convection. In the next section, we derive a low-dimensional dynamical system to reconcile the observations from DNS.

\section{Low-dimensional dynamical system}
\label{sec-lowdim}

In this section, we derive a low-dimensional ODE system for thermal convection in an annulus. Importantly, the system is cast in terms of physically relevant variables: the average angular momentum $L(t)$ and CoM coordinates $(\xc(t),\yc(t))$ as defined in \cref{Ldef,COM} respectively. These variables permit a transparent description of the circulatory flow fields and buoyancy variations characterizing thermal convection. 

The ODE system derives systematically from Galerkin truncation of the governing equations, a strategy that has been successfully employed for other systems \cite{majda2019statistical, moore2020anomalous, sun2023normal, sun2023parameter}. In particular, we perform a Fourier-Laurent expansion of solutions to the NSB system and truncate to the lowest-order capable of satisfying all boundary conditions on the inner and outer rings of the annulus. The velocity field that results from this process is a shear flow in the angular direction, $u=u(r)$ and $v=0$, which best approximates the true flow when the annulus is relatively narrow, i.e.~$r_0 \approx 1/2$. For this reason, we set $r_0 = 0.4$ in the majority of numerical experiments. For smaller values of $r_0$, we expect the ODE model to still capture the leading-order dynamics of the full system, but with larger quantitative differences.

\subsection{Derivation of ODE system}

When expressed in polar coordinates, the angular component of \cref{NS} and the incompressibility constraint \cref{incomp} take the form
\begin{align}
\label{NSeth}
& u_t + v u_r + \frac{1}{r} u u_\theta + \frac{1}{r} u v = -\frac{1}{r} p_\theta + \Ra \Pra \, T \cos \theta + \Pra \left( u_{rr} + \frac{1}{r} u_r + \frac{1}{r^2} u_{\theta \theta} - \frac{1}{r^2} u + \frac{2}{r^2} v_\theta \right) \, , \\
\label{incomp2}
& v_r + \frac{1}{r} v + \frac{1}{r} u_\theta = 0 \, .
\end{align}
Multiplying \cref{NSeth} by $r^2$, integrating over the annulus $\Omega$, applying incompressibility \cref{incomp2} and the no-slip condition \cref{noslip}, and using the CoM definition \cref{COM} gives the evolution equation
\begin{equation}
\label{Ldot1}
\dot{L} = -\Ra\, \Pra \, \xc + 
\frac{\Pra}{\Area} \int_0^{2\pi} \left(r^2 u_r \right)\Big|_{r_0}^{r_1} \, d\theta.
\end{equation}
This equation is {\em exact} within the NSB framework.
The first term, $ -\Ra\, \Pra \, \xc $, represents torque due to an off-center CoM, much like in a pendulum system. This torque tends to destabilize the top-heavy conductive state. For example, if the CoM is perturbed to the left, $\xc<0$, then $ -\Ra \Pra \, \xc > 0$, which increases $L$. Since the conductive CoM is raised, $y_0 >0$, increasing CCW angular momentum carries $\xc$ farther left, creating the positive feedback associated with instability. 
The second term involving $ \left(r^2 u_r \right)\big|_{r_0}^{r_1}$ acts as damping. For example, if the rotation is primarily CCW with $L>0$, then $u > 0$ on average. Thus, in order to satisfy the no-slip boundary conditions, $u$ must decrease as $r$ approaches the inner or outer boundary, $r\to r_0^{+}$ or $r \to 1/2^{-}$ respectively, both giving $ \left(r^2 u_r \right) \big|_{r_0}^{r_1} <0$ and thus reducing the angular momentum.

We next introduce some approximations to supplement the exact evolution \cref{Ldot1} and obtain a closed system for the variables $L(t), \xc(t), \yc(t)$.
Since the temperature distribution $T(r,\theta,t)$ is periodic in $\theta$, it can be written as a Fourier series with no approximation made,
\begin{equation}
\label{TFourier}
T(r,\theta,t) = a_0(r,t) + \sum_{n=1}^{\infty}  a_n(r,t) \cos n\theta + b_n(r,t) \sin n\theta ,
\end{equation}
From \eqref{Tinner}--\eqref{Touter}, the coefficients inherit boundary conditions
\begin{align}
\label{abinner}
& \pdi{a_n}{r} = \pdi{b_n}{r} = 0 \quad \text{ at } r=r_0, \\
\label{abouter}
& a_0 = 1/2, b_1 = -1/2, \text{  all others vanish at } r=1/2.
\end{align}
Similarly, both velocity components are periodic in $\theta$, and so each can be written as a (complex) Fourier series
\begin{align}
\label{uvFourier}
u(r,\theta,t) = \sum_{n=-\infty}^{\infty} \hat{u}_n(r,t) e^{i n \theta},  \quad
v(r,\theta,t) = \sum_{n=-\infty}^{\infty} \hat{v}_n(r,t) e^{i n \theta} .
\end{align}
The no-slip boundary conditions, \cref{noslip}, and incompressibility, \cref{incomp2}, respectively yield the conditions
\begin{align}
\label{uvh_bc}
&\hat{u}_n(r,t)=\hat{v}_n(r,t) = 0
\hspace{25pt} \text{at } r=r_0 \text{ and } r=1/2, \\
\label{uvh_incomp}
&i n \hat{u}_n + \hat{v}_n + r \pdi{\hat{v}_n}{r} = 0
\end{align}
holding for each $n$.

We now aim to truncate the Fourier expansions, \cref{TFourier,uvFourier}, {\em to the lowest order capable of satisfying all boundary conditions}. In particular, we retain up to the $n=1$ mode in the temperature field and the $n=0$ in the velocity field. It is necessary to retain the $n=1$ mode in the temperature field to satisfy the thermal condition, \cref{abouter}, whereas only the $n=0$ mode in the flow field is needed to satisfy the no-slip conditions, \cref{uvh_bc}. Enforcing incompressibility, \cref{uvh_incomp}, implies that $\hat{v}_0(r,t) = 0$, which shows that the leading-order flow structure is shear $(u,v) \sim (\hat{u}_0(r,t), 0)$.

The thermal transport equation, \cref{trans}, written in polar coordinates is
\begin{equation}
\label{HeatTrans}
T_t + \frac{u}{r} T_\theta + v T_r = \frac{1}{r} \pd{}{r} \left( r T_r \right) + \frac{1}{r^2} T_{\theta \theta} .
\end{equation}
Inserting the Fourier expansion \cref{TFourier} and the truncated velocity fields, $(u,v) = (\hat{u}_0(r,t), 0)$, into \cref{HeatTrans}, multiplying by $r^2$, and projecting onto Fourier mode $n$ gives
\begin{align}
\label{andot}
& r^2 \dot{a}_n = -n r \, \hat{u}_0(r,t) \, b_n - n^2 a_n + r \pdi{}{r} \left( r \pdi{}{r} a_n \right) , \\
\label{bndot}
& r^2 \dot{b}_n = +n r \, \hat{u}_0(r,t) \, a_n - n^2 b_n + r \pdi{}{r} \left( r \pdi{}{r} b_n \right) .
\end{align}
At order $n=0$, the above gives a diffusion equation for $a_0(r,t)$, 
\begin{equation}
\label{a0}
\dot{a}_0 = r^{-1} \pdi{}{r} \left( r \pdi{}{r} a_0 \right) .
\end{equation}
Boundary conditions \eqref{abinner}--\eqref{abouter} imply $\lim_{t\to \infty} a_0(r,t) = 1/2$, regardless of initial conditions. We will therefore set $a_0 = 1/2$ henceforth, as variations from this value simply represent transient dynamics that are decoupled from the rest of the system.

From \cref{COM}, the CoM coordinates are given by
\begin{align}
\label{COM_int}
\xc(t) = -\frac{\pi}{\Area} \int_{r_0}^{1/2} r^2 a_1(r,t) \, dr , \qquad
\yc(t) = -\frac{\pi}{\Area} \int_{r_0}^{1/2} r^2 b_1(r,t) \, dr .
\end{align}
Differentiating with respect to time, inserting \cref{andot,bndot} with $n=1$, and simplifying yields the formulas
\begin{align}
\label{xcdot}
& \dot{\xc} = \frac{\pi}{\Area} \int_{r_0}^{1/2} r \hat{u}_0(r,t) b_1(r,t) \, dr
- \frac{\pi}{\Area}  \left(r^2 \pd{a_1}{r} - r a_1 \right) \Big|_{r_0}^{r_1}, \\
\label{ycdot}
& \dot{\yc} = - \frac{\pi}{\Area}  \int_{r_0}^{1/2} r \hat{u}_0(r,t) a_1(r,t) \, dr
- \frac{\pi}{\Area}  \left(r^2 \pd{b_1}{r} - r b_1 \right) \Big|_{r_0}^{r_1}.
\end{align}

We now assume special forms for the radial dependence of the variables $a_1(r,t)$, $b_1(r,t)$, and $\hat{u}_0(r,t)$. Guided by the conductive-state solution, \eqref{conductive}, we assume truncated Laurent expansions for the coefficients $a_1$ and $b_1$:
\begin{align}
\label{a1}
& a_1(r,t) =  \frac{1}{2} A(t) (2r-1) \left(1 - 2 r_0^2 r^{-1} \right), \\
\label{b1}
& b_1(r,t) = -\frac{1}{2} + \frac{1}{2} B(t) (2r-1) \left(1 - 2 r_0^2 r^{-1} \right).
\end{align}
These are the most general Laurent expansions containing powers $r^{-1}, r^{0}$, $r^{1}$ and satisying boundary conditions \cref{abinner,abouter}.
Setting $A(t) = 0$, $B(t) = -(4r_0^2+1)^{-1}$ recovers the conductive-state solution, \cref{conductive}, exactly, whereas allowing these coefficients to vary creates different buoyancy fields.

Also guided by the Laurent expansion, we assume the following form for the angular velocity
\begin{equation}
\label{uu}
\hat{u}_0(r,t) = C(t) (r-r_0) \left(1-2r \right) r^{-1}.
\end{equation}
Similarly, this is the most general Laurent expansion that contains powers $r^{-1}$, $r^{0}$, $r^{1}$ and that satisfies the no-slip  conditions \cref{uvh_bc}. Setting $C(t)=0$ trivially recovers the conductive state, whereas allowing this coefficient to vary creates different circulatory flow fields.

Inserting \cref{a1,b1} into \cref{COM_int} and integrating yields the following linear relationships between the CoM coordinates and the coefficients $A(t), B(t)$:
\begin{align}
\label{COMAB1}
& \xc(t) = \left( \frac{(1-2r_0)^2 (1 + 6r_0 + 16r_0^2)}{48(1+2 r_0)} \right) \, A(t), \\
\label{COMAB2}
& \yc(t) =  \frac{1 + 2r_0 + 4 r_0^2}{12(1+2r_0)}
+ \left( \frac{(1-2r_0)^2 (1 + 6r_0 + 16r_0^2)}{48(1+2 r_0)} \right) \, B(t) .
\end{align}
Meanwhile, from \cref{uu} and the definition of angular momentum, \cref{Ldef}, $L(t)$ relates linearly to $C(t)$ through
\begin{align}
\label{L_int}
L(t) = \frac{2\pi}{\Area} \int_{r_0}^{1/2} r^2 u(r,t) \, dr =
\frac{(1-2 r_0)^2}{12} \, C(t) .
\end{align}

Inserting \cref{uu} into \cref{Ldot1}, using the linear relationship \cref{L_int}, and simplifying gives the evolution equation
\begin{align}
\label{Ldot2}
\dot{L} = - \Ra \Pra \, \xc - \alpha \Pra \, L ,
\end{align}
where $\alpha = 48/(1-2r_0)^2$. Meanwhile, inserting Eqs.~\eqref{a1}--\eqref{L_int} into \cref{xcdot,ycdot}, performing exact integration and simplifying gives the following evolution equations for the CoM coordinates:
\begin{align}
\label{xdot}
&\dot{\xc} = -k L \yc + \gamma L - \beta \xc , \\
\label{ydot}
&\dot{\yc} = +k L \xc + \delta - \beta \yc .
\end{align}
Here, the coefficients $\alpha, \beta, \delta, k$, and $\gamma$ are each functions of $r_0$ only, as given by:
\begin{align}
\label{alpha}
&\alpha = \frac{48}{(1-2r_0)^2}, \quad
\beta = \frac{48 (1+4r_0^2)}{(1-2r_0)^2 (1+6r_0+16r_0^2)}, \quad
\delta =  \frac{3 (1+12r_0^2)}{(1-2r_0)^2 (1+6r_0+16r_0^2)}, \\[5pt]
\label{keq}
&k = 24 \frac{(1-2r_0) 
(1 -6r_0 -4r_0^2 -88r_0^3 + 32r_0^4) 
-96 r_0^3 \ln{(2r_0)}} {(1-2r_0)^5 (1+6r_0+16r_0^2)}, \\[5pt]
\label{gamma}
&\gamma = \frac{(1-4r_0^2)
(1 -8r_0 -224r_0^3 -80r_0^4) - 192r_0^3(1+2r_0+4r_0^2)\ln{(2r_0)}}{ (1-2r_0)^5 (1+2r_0)(1+6r_0+16r_0^2)}.
\end{align}

Two important length scales naturally arise from grouping like terms in \cref{xdot,ydot}:
\begin{align}
\label{y0}
& y_0 = {\delta}/{\beta} && \text{The height of the conductive-state CoM} , \\
\label{y1}
& y_1 = {\gamma}/{k} && \text{The height of the pendulum fulcrum} ,
\end{align}
where the interpretations will be justified  momentarily. With these definitions, the self-contained dynamical system \cref{Ldot2,xdot,ydot} becomes
\begin{align}
\label{Ldot*}
&\dot{L} = - \Ra \Pra \, \xc - \alpha \Pra \, L , \\
\label{xdot*}
&\dot{\xc} = -k L (\yc - y_1) - \beta \xc ,  \\
\label{ydot*}
&\dot{\yc} = +k L \xc - \beta (\yc - y_0) .
\end{align}

The above form offers some important physical insight. First, if there is no flow $L=0$, \cref{xdot*,ydot*} show that the CoM converges to the point $(\xc, \yc) = (0, y_0)$ with $y_0$ given by \cref{y0}. At the same time, no fluid motion produces the conductive-state solution \cref{conductive}, with CoM height given by \cref{y0_cond}. Hence, \cref{y0} must correspond to \cref{y0_cond}, and this can be verified directly; both give equivalent formulas for the height of the conductive-state CoM. 

\begin{figure}[htb]
 \includegraphics[width=6in]{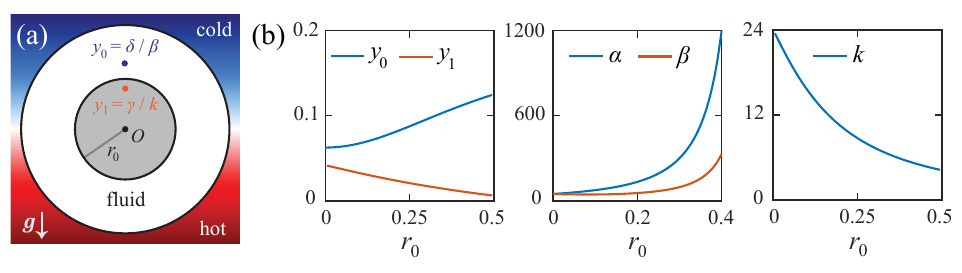}
  \caption{Pendulum structure of the ODE system. (a) \Cref{Ldot*,xdot*,ydot*} correspond to a pendulum with fulcrum $y_1$ and forcing term that drives the CoM towards $y_0$. (b) The constants in the ODE model depend on $r_0$ only; $y_0$ always lies above the fulcrum, $y_1$, implying that $\dy = y_0-y_1>0$ for any $r_0$.}
\label{fig4}
\end{figure}

Second, if $\beta=0$, \cref{Ldot*,xdot*,ydot*} are {\em mathematically identical} to those of a damped pendulum with angle $\phi(t)$, mass $m$, and length $l$. \Cref{Ldot*,xdot*,ydot*} are simply written in terms of the bob's Cartesian coordinates $(\xc, \yc) = (l \sin \phi, \, y_1 - l \cos \phi)$ and angular momentum $L = m l^2 \dot{\phi}$. In this analogy, the gravitational constant is $g = k l^2 \Ra \Pra$, and the damping coefficient is $\alpha \Pra$. Most importantly, \cref{xdot*} shows that fulcrum of the pendulum lies at the point $(0,y_1)$ with $y_1$ given by \cref{y1}.

The terms in \cref{xdot*,ydot*} with prefactor $\beta$ arise from the interaction of boundary heating and thermal diffusion. These terms drive the CoM towards the conductive-state CoM $(0,y_0)$. Through \cref{y0,y1}, it can be shown that $0 < y_1 < y_0$ for any $r_0$, implying that these terms act to raise the CoM above the fulcrum and, hence, tend to destabilize the system. \Cref{fig4}(a) illustrates these two heights, $y_0$ and $y_1$, and \cref{fig4}(b) shows their dependence on $r_0$. A crucial parameter that appears in the stability analysis is the distance $\dy = y_0 - y_1 > 0$, which is positive for any $r_0$ as seen in \cref{fig4}(b). Also seen in \cref{fig4}(b) is the dependence of the parameters $\alpha, \beta$, and $k$ on $r_0$. 

It is important to remember that the constants in the ODE model, $\alpha, \beta, k, y_0, $ and $y_1$, are purely geometric in that they depend  on $r_0$ only as given by \cref{alpha,keq,gamma}. The only parameters that depend on other physical properties, such as the strength of thermal forcing, the viscosity, etc., are $\Ra$ and $\Pra$.

\subsection{Simulation of ODE system in comparison to DNS}

\begin{figure}[htb!]
 \includegraphics[width=6.5in]{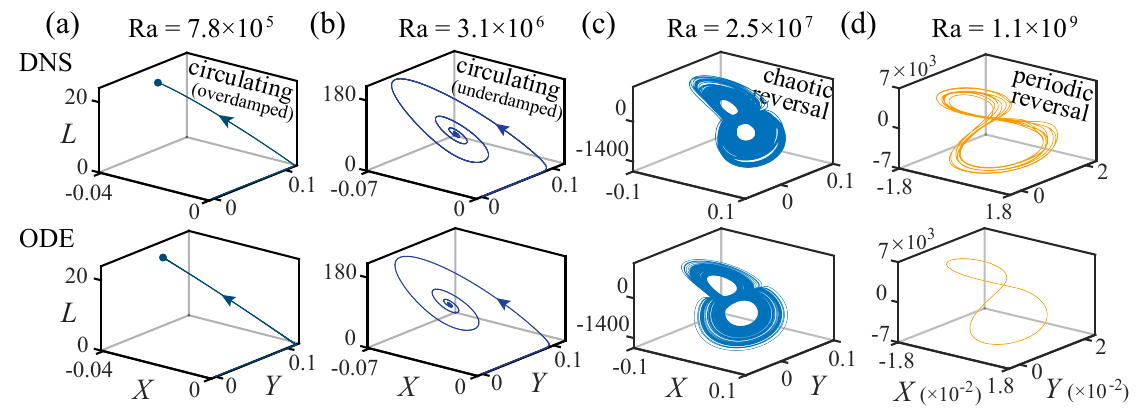}
  \caption{Comparison between ODE model \cref{Ldot*,xdot*,ydot*} and fully-resolved DNS. (a) At $\Ra = 7.8 \times 10^5$, the dynamical system is overdamped and the solution trajectory ($L$, $X$, $Y$) quickly approaches the steady circulating state. (b) At $\Ra = 3.1 \times 10^6$, the dynamical system becomes underdamped and the solution is an inward spiral towards the steady circulating state. (d) Increasing  $\Ra{}$ to  $2.5\times10^7$ brings the trajectory to chaos, whose shape indicates a strange attractor with fractal dimension of 1.4. (e) Limit cycle appears at $\Ra{} = 1.1\times10^9$, indicating a periodic solution. In all simulations, $\Pra = 4$ and $r_0 = 0.4$. Movies of (b)-(d) are included in the \si{}.}
\label{fig5}
\end{figure}

In this section, we discuss numerical solutions of the ODE system, \cref{Ldot*,xdot*,ydot*}, in comparison to the fully-resolved DNS of  \cref{NS,trans,incomp,RaPra,noslip,Tinner,Touter}. Numerical solutions of the ODE system are found with \textsc{Matlab}'s \textit{ode45}. As in previous sections, we fix $\Pra = 4$ and $r_0 = 0.4$, and vary the Rayleigh number. \Cref{fig5} shows solution trajectories of $(L(t), \xc(t), \yc(t))$ computed from both the DNS (top panel) and the ODE system (bottom panel) for a sequence of four Rayleigh numbers. In each case, we prescribe the same initial conditions in the DNS and ODE system. The resulting solution trajectories are remarkable similar in all four cases, and in fact nearly identical in the first two [\cref{fig5}(a)--(b)], suggesting that the simplified ODE system recovers detailed convective dynamics across a range of Rayleigh numbers.

The first two cases [\cref{fig5}(a)--(b)] feature the lowest Rayleigh numbers, $\Ra = 7.8 \times 10^5$ and $3.1 \times 10^6$ respectively. In each case, the solution $(L(t), \xc(t), \yc(t))$ converges to a fixed point with non-zero angular momentum, $L\ne0$, and a CoM that is raised, $\yc >0$, and off-set, $\xc \ne 0$. This type of fixed point corresponds to the steadily circulating state seen in \cref{fig1}(c); the fluid rotates in either the CW or CCW direction at constant rate. The cases shown in \cref{fig5}(a)--(b) exhibit CCW rotation $L>0$ as a result of the initialization. The main difference between \cref{fig5}(a) and \cref{fig5}(b), is that at  $\Ra = 7.8 \times 10^5$ the system converges to the circulating state as an overdamped oscillator, and at higher Rayleigh, $\Ra = 3.1 \times 10^6$, the system converges as an underdamped oscillator.

At $\Ra = 2.5 \times 10^7$, \cref{fig5}(c) shows more complex solution trajectories that appear to fill a higher-dimensional set. Measurements in Section \ref{sec-bir} of this set's fractal dimension yield a value of 1.4, characteristic of a strange attractor. Each sign change of $L$ seen in \cref{fig5}(c) indicates a reversal of the LSC. The view of the $L(t)$ time-series from \cref{fig2} shows  these reversals to occur erratically, suggesting chaotic dynamics. While it is difficult to directly compare two chaotic time series, the attracting sets obtained from DNS and from the ODE system appear remarkably similar, suggesting that the reduced ODE model captures the main features of convective dynamics in this regime.

\Cref{fig5}(d) shows the highest Rayleigh number, $\Ra = 1.1 \times 10^9$. In this case, the ODE dynamics converge to a stable limit cycle, seen as the simple, closed curve in the bottom panel (the early-time convergence to the limit cycle is not shown). Since $L$ changes sign, this limit cycle corresponds to periodic reversals of the LSC, much like was observed in the DNS at high $\Ra$ (see the right-most panel of \cref{fig2}). The top panel of \cref{fig5}(d) shows that the DNS dynamics approximately follow a similarly shaped limit cycle, though with fluctuations about the cycle. In the DNS, the LSC reversals are nearly periodic, but with a small amount of variability. This variability will be quantified further in Section \ref{sec-pendulum} by measurements of the frequency spectrum. Despite the presence of fluctuations, the main shape of the limit cycle from DNS and the ODE model in \cref{fig5}(d) appear similar, again suggesting that the ODE model captures the leading-order convective dynamics well.

The close match between the ODE model and DNS observed in \cref{fig5} suggests that the ODE model successfully captures the main features of convective dynamics across a range of Rayleigh numbers. Since the ODE model is substantially simpler, it may offer insight into the transitions between the different convective steps, and this idea is explored in the next section.

\section{Stability and bifurcation analysis}
\label{sec-bir}

To better understand the transitions between different convective states, we now examine the stability and bifurcations of the ODE model, \cref{Ldot*,xdot*,ydot*}. First, to locate the fixed points, we set $\dot{L}=0$ and $\dot{\yc}=0$ to obtain
\begin{align}
\xc = -\frac{\alpha}{\Ra} L , \qquad
\yc = y_0 - \frac{k\alpha}{\beta \Ra} L^2 .
\end{align}
Then setting $\dot{\xc}=0$ yields a condition for the fixed points in terms of $L$ only,
\begin{equation}
L \left[ k^2 \alpha L^2 - \left( \beta k \Ra \dy - \alpha \beta^2 \right) \right] = 0 .
\end{equation}
There can be up to three roots of this cubic equation:
\begin{align}
\label{fpts}
L = 0 , \qquad
L = \pm L_1 = \pm \frac{\beta}{k} \sqrt{ \frac{k \Ra}{\alpha \beta} \dy -1 } .
\end{align}
The first root, $L=0$, corresponds to the conductive state. The second two roots, $L = \pm L_1$, are real only if the term under the radical is positive. These roots correspond to circulating states of constant angular momentum in either the CCW or CW direction ($+L_1$ and $-L_1$ respectively). The three roots yield three possible fixed points of the system:
\begin{align}
\label{CondState}
L &=0, \quad \xc = 0, \quad \yc = y_0 
&&\mbox{The conductive state},\\
\label{CircState}
L &= \pm L_1, \quad \xc = \mp\frac{\alpha}{\Ra} L_1, \quad  \yc = y_1 + \frac{\alpha\beta}{k \Ra} 
&&\mbox{The circulating states}.
\end{align}
For general $(L,\xc,\yc)$, the Jacobian of \cref{Ldot*,xdot*,ydot*} is given by
\begin{equation}
\label{Jacobian}
J(L,\xc,\yc) = \left[ {\begin{array}{ccc}
   -\alpha\,\Pra& -\Ra\,\Pra & 0 \\
   -k(\yc-y_1)  & -\beta    & -kL \\
   k\xc         & kL    & -\beta
  \end{array} } \right].
\end{equation}
Evaluating the Jacobian determines the type and stability of each fixed point.

\subsection{$\Ra < \Ra^*_1$, stable conductive state}

We now analyze bifurcations with respect to increasing Rayleigh number, $\Ra$, while holding $r_0$ and $\Pra$ fixed. In particular, for Rayleigh numbers below the critical value,
\begin{equation}
\label{ra1}
\Ra_1^*  = \frac{\alpha \beta}{k \dy},
\end{equation}
the term under the radical in \cref{fpts} is negative and so the circulating-state fixed points do not exist.
Thus, for $\Ra < \Ra_1^*$ the conductive state, \cref{CondState}, is the only fixed point of the system. Evaluating the Jacobian gives
\begin{equation}
J_{\text{cond}} = \left[ {\begin{array}{ccc}
   -\alpha\,\Pra & -\Ra\,\Pra & 0 \\
   -k \dy & -\beta & 0 \\
   0 & 0 & -\beta
  \end{array} } \right].
\end{equation}
Consider the three eigenvalues $z_1, z_2, z_3$ of this matrix. Due to the zeros in the last row and last column, one eigenvalue is $z_1 = -\beta$. The other two are eigenvalues of the smaller $2 \times 2$ subsystem that excludes the final row and final column. The trace of this subsystem is negative, and the determinant is equal to $ \Pra (\alpha \beta - k \Ra \dy) $, which transitions from positive to negative precisely as $\Ra$ crosses the threshold $\Ra_1^{*}$.
Thus, for $\Ra < \Ra_1^{*}$, all three eigenvalues are negative and so the conductive state corresponds to a stable node. For $\Ra > \Ra_1^{*}$, two eigenvalues are negative and one is positive, meaning the conductive state is a saddle point.

\subsection{$\Ra_1^*\leq\Ra < \Ra_2^*$, bistable circulating states}

As $\Ra$ crosses the critical value $\Ra_1^*$, the two circulating-state fixed points emerge, and, simultaneously, the conductive state loses stability. That is, a supercritical pitchfork bifurcation occurs. Evalauting the Jacobian at the circulating fixed points will show that they emerge as stable fixed points and then undergo stability transitions at yet higher Rayleigh numbers.

In particular, the Jacobian matrix, \cref{Jacobian}, evaluated at each circulating fixed-point, \cref{CircState}, is given by
\begin{equation}
J_\pm = \left[ {\begin{array}{ccc}
   -\alpha\, \Pra & -\Ra\, \Pra & 0 \\
   -\alpha\beta/\Ra & -\beta & {\mp k L_1} \\
   {\mp \alpha k L_1/\Ra} & {\pm k L_1} & -\beta
  \end{array} } \right].
\end{equation}
The characteristic polynomial of this matrix is
\begin{align}
\label{character}
P(z) &= z^3+c_2z^2 +c_1z+c_0 , \\
\label{c0} c_0 &= 2\alpha k^2 L_1^2 \,\Pra , \\
\label{c1} c_1 &= k^2 L_1^2 +\beta^2+\alpha\beta\,\Pra , \\
\label{c2} c_2 &= \alpha\,\Pra+2\beta .
\end{align}
This cubic polynomial has three roots, $z_1, z_2$, and $z_3$. At least one root is guaranteed to be real, while the other two may either be real or form a complex-conjugate pair.
The discriminant $\Delta$ determines which occurs,
\begin{equation}
\label{disc}
\Delta = (z_1 - z_2)^2(z_2-z_3)^2(z_1-z_3)^2 =
c_1^2 c_2^2 - 4c_0 c_2^3  + 18c_0 c_1 c_2 -4 c_1^3-27c_0^2.
\end{equation}
If $\Delta \ge 0$, then all three roots are real, whereas if $\Delta < 0$, then two of the roots are complex conjugates.

We first consider $\Ra$ slightly above the critical value $\Ra_1^*$, in which case $L_1^2 = O(\Ra - \Ra_1^*)$ is small. Substituting into \eqref{disc} gives
\begin{equation}
\Delta = c_1^2 \alpha^2 {\Pra}^2 + O(\Ra - \Ra_1^*).
\end{equation}
Thus, if $\Ra > \Ra_1^*$ and $\Ra$ is sufficiently close to $\Ra_1^*$, then $\Delta > 0$ and so all three roots are real. Furthermore, the coefficients $c_0, c_1, c_2$ are all positive, and so there cannot be any positive roots of \cref{character}. Therefore, in this case of $\Ra$ slightly above $\Ra_1^{*}$,  all three eigenvalues are negative, and so the circulating states correspond to bistable nodes.

As $\Ra$ continues to increase, the discriminant eventually becomes negative implying that two eigenvalues become complex. The precise Rayleigh number at which this occurs, denoted $\Ra_{3/2}^*$, can be determined by setting $\Delta = 0$ in \eqref{disc}. As $\Ra$ crosses $\Ra_{3/2}^*$, the circulating states transition from stable nodes to stable spirals. In the former stage, the system behaves as an overdamped oscillator as seen in \cref{fig5}(a), and in the later stage, as an underdamped oscillator as seen in \cref{fig5}(b).

As $\Ra$ increases further beyond $\Ra_{3/2}^*$, the circulating states eventually lose stability. To determine where the transition occurs, we use Vieta's formulas
\begin{align}
\label{Vieta1}
z_1 + z_2 + z_3 &= -c_2 , \\
\label{Vieta2}
z_1 z_2 +z_2 z_3 + z_1 z_3 &= c_1 , \\
\label{Vieta3}
z_1 z_2 z_3 &= -c_0 .
\end{align}
where $z_1, z_2, z_3$ are the three eigenvalues, and $c_0, c_1, c_2$ are the coefficients given by \cref{c2,c1,c0}. Let $z_1$ denote the real eigenvalue and $z_{2,3} = \sigma \pm i \omega$ the complex-conjugate pair. Then the circulating-state fixed points are stable spirals if $\sigma<0$ and unstable spirals if $\sigma>0$. The transition occurs at $\sigma = 0$, which implies that $z_2 + z_3 = 0$ and $z_2 z_3 = \omega^2$. Inserting into \cref{Vieta1,Vieta2,Vieta3} gives $z_1 = -c_2$, $\omega^2 = c_1$, and $z_1 \omega^2 = -c_0$ respectively, which combine to give $c_0 = c_1 c_2$. Substituting this relationship into the definition of the coefficients, \cref{c2,c1,c0}, yields
\begin{equation}
\label{alg}
2\alpha k^2 L_1^2 \,\Pra= (k^2 L_1^2 +\beta^2+\alpha\beta\,\Pra) (\alpha \,\Pra+ 2\beta),
\end{equation}
Recall that the Rayleigh number appears in $k^2 L_1^2$ through 
\begin{equation}
k^2 L_1^2 = \alpha^{-1} \beta k \Ra \dy - \beta^2.
\end{equation}
Therefore, solving \eqref{alg} for $k^2 L_1^2$ yields the critical Rayleigh number $\Ra_2^*$ at which the circulating states lose stability, 
\begin{equation}
\label{ra2}
\Ra_2^* = \frac{\alpha^2 \, \Pra}{k \dy} \left( \frac{\alpha \Pra + 4\beta}{\alpha \Pra - 2\beta} \right).
\end{equation}
Beyond this threshold, all fixed points of the system are unstable.

\subsection{$\Ra \geq \Ra_2^*$, chaos and eventual return to order}

As $\Ra$ crosses $\Ra_2^*$, a Hopf bifurcation occurs and the circulating states change from stable to unstable spiral points. Past this critical value, all fixed points of the system are unstable, thereby introducing the possibility of chaotic dynamics as supported by the numerical observations in \cref{fig5}(c). Before analyzing the chaotic state, the explicit form of \cref{ra2} offers a few simple observations. First, for $\Pra$ smaller than the critical value
\begin{equation}
\label{Pra*}
\Pra^*= 2\beta/\alpha,
\end{equation}
the denominator in \cref{ra2} is negative, implying that there is no $\Ra_2^*$ threshold. That is,  if $\Pra \le \Pra^*$ and the circulating states exist, then they remain stable no matter how large the Rayleigh number is. Thus, $\Pra \le \Pra^*$ precludes the possibility of chaos; trajectories are simply attracted to one of the bistable circulating states or to the conductive state.

The second observation is that taking the limit $\Pra \to \infty$ in \cref{ra2} shows that the threshold $\Ra_2^*$ scales linearly with $\Pr$ with prefactor $\alpha^2/(k\dy)$. That is, large $\Pra$ numbers require large $\Ra$ values to reach the chaotic regime. Since large $\Ra$ values generally require greater computational expense in DNS, this observation suggests that the most practical way to realize the chaotic state in the DNS is to choose $\Pr$ above the threshold \cref{Pra*}, but not too large.

The value $r_0=0.4$ chosen for \cref{fig5} gives $\alpha = 1200, \beta = 330, k = 5.5, \dy =0.1$, $\Pra^* = 0.55$, and $\Ra_1^* = 7.3 \times 10^5$. These values, combined with the choice $\Pra=4$, yields $\Ra_2^* = 1.6 \times 10^7$. Thus, the values of $\Ra = 7.8 \times 10^5$ and $\Ra = 3.1 \times 10^6$ used in \cref{fig5}(a)--(b) lie in the range $[\Ra_1^*, \Ra_2^*]$, for which stability analysis predicts the circulating states to be stable fixed points. This prediction is confirmed by both the DNS and ODE numerical trajectories shown in the figure. Meanwhile, the value $\Ra = 2.5 \times 10^7$ used in \cref{fig5}(c) lies above the $\Ra_2^*$ threshold for which the analysis predicts all fixed points to be unstable. Again, this prediction is consistent with the chaotic numerical trajectories observed in the figure. Lastly, the value $\Ra = 1.1 \times 10^9$ also exceeds the $\Ra_2^*$ threshold, but rather than chaotic dynamics, trajectories converge towards a stable limit cycle. In summary, all of the numerical trajectories from both DNS and the ODE system shown in \cref{fig5} are consistent with the threshold values, \cref{ra1,ra2}, predicted by stability analysis.


An important insight provided by the numerical trajectories in \cref{fig5}(c)--(d) is that, for $\Ra \ge \Ra_2^*$, the long-time dynamics may either be chaotic, as in \cref{fig5}(c), or periodic, as in \cref{fig5}(d); both behaviors are consistent with the conclusion from stability analysis that all fixed points are unstable. \Cref{fig5}, however, only shows a selection of four particular Rayleigh numbers, and thus offers only a coarse evaluation of the predictions from stability analysis. In the next section, we conduct a more thorough comparison.

\subsection{Stability results in comparison to numerical trajectories}

\begin{figure}[htb]
 \includegraphics[width=6in]{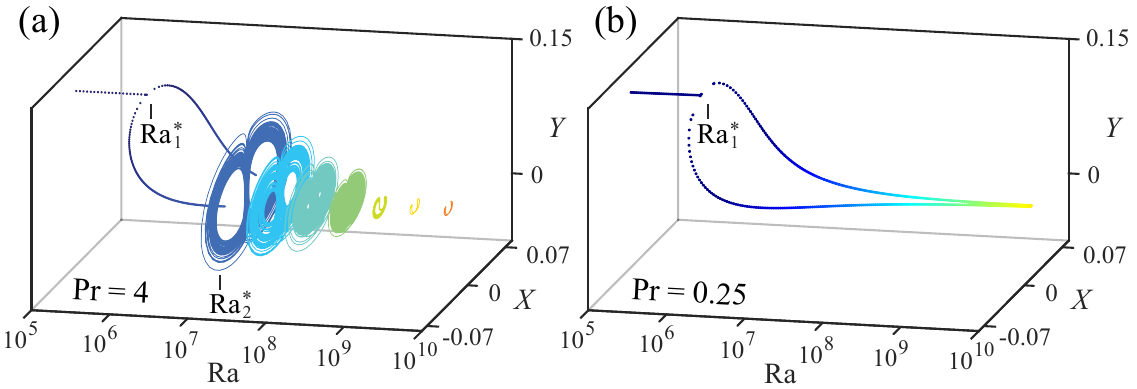}
  \caption{Bifurcation diagrams show long-time trajectories of CoM coordinates, $(\xc,\yc)$, computed numerically from the ODE system \cref{Ldot*,xdot*,ydot*}. Trajectories are plotted against Rayleigh number, with $r_0 = 0.4$ and with $\Pra = 4$ and $0.25$ in (a) and (b) respectively. In both cases, a pitchfork bifurcation occurs at $\Ra_1^*$ predicted by \cref{ra1}, where the bistable circulating states emerge as stable nodes. (a) For $\Pra = 4$, a second, Hopf bifurcation occurs at $\Ra_2^*$ predicted by \cref{ra2}, where the circulating states lose stability and chaotic dynamics emerge. For $\Ra > 10^9$, chaos gives way to periodic dynamics. (b) For $\Pra = 0.25$, below the threshold value $\Pra^* = 0.55$ from \cref{Pra*}, the circulating states remain stable for arbitrarily large $\Ra$.}
\label{fig6}
\end{figure}

In this section, we systematically compare the predictions of the stability analysis with the numerical trajectories of \cref{Ldot*,xdot*,ydot*}. In particular, we examine bifurcations with respect to increasing Rayleigh number. \Cref{fig6} shows long-time numerical trajectories of the CoM coordinates, $(\xc(t), \yc(t))$, plotted against $\Ra$ on the horizontal axis ($L(t)$ is not shown). The Prandtl number is set to $\Pra = 4$ and 0.25 in \cref{fig6}(a) and (b) respectively. 

For low $\Ra$, both figures show that long-time dynamics collapse to a single stable fixed point that corresponds to the conducting state. As $\Ra$ increases, a supercritical pitchfork bifurcation occurs precisely at the value $\Ra = \Ra_1^*$ predicted by \cref{ra1}. This value, $\Ra_1^*= 7.3 \times 10^5$, is independent of $\Pra$, and thus the pitchfork bifurcation occurs at exactly the same location in both (a) and (b). The two branches to the right of the pitchfork represent the bistable circulating states. 

As $\Ra$ increases further, \cref{fig6}(a) shows that a Hopf bifurcation occurs at the value $\Ra = \Ra_2^* = 1.6 \times 10^7$ predicted by \cref{ra2}. Here, the circulating states lose stability and give way to chaotic dynamics, seen by the blue, turquoise, teal, and green trajectories. Meanwhile, \cref{fig6}(b) does not exhibit a Hopf bifurcation. In \cref{fig6}(b), the Prandtl number $0.25$ lies below the critical value $\Pra^* = 0.55$ predicted by \cref{Pra*}. Therefore, the bistable circulating states remain stable for arbitrarily large $\Ra$, as is consistent with the trajectories seen in \cref{fig6}(b).

Returning to \cref{fig6}(a), as $\Ra$ increases further beyond $\Ra_2^*$, the chaotic dynamics eventually subside and give way to the more confined and regular dynamics, shown by the green, yellow, and red trajectories. The figure suggests the transition to occur at roughly $\Ra = 10^9$. These trajectories resemble small circular arcs, consistent with pendulum motion. Furthermore, \cref{fig5}(d) shows that the corresponding dynamics are periodic, or in the case of DNS, nearly periodic.

\begin{figure}[htb]
\includegraphics[width=4.5in]{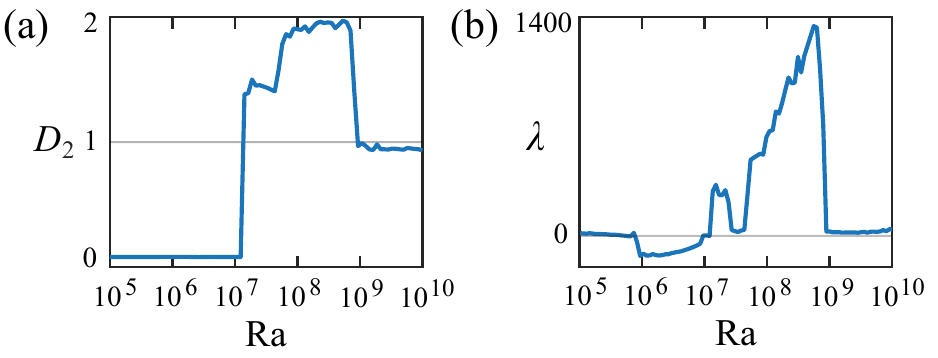}
\caption{The fractal dimension $D_2$ and Lyapunov exponent $\lambda$ from trajectories in \cref{fig6}(a) characterize states of order and chaos. (a) For $\Ra < \Ra_2^*$, the fractal dimension is nearly zero, reflecting convergence to a fixed point. As $\Ra$ crosses $\Ra_2^*$, the fractal dimension abruptly increases to a value in the range $1<D_2<2$, characteristic of a strange attractor. For $\Ra > 10^9$, $D_2$ drops back to unity, reflecting convergence to a one-dimensional limit cycle. (b) For $\Ra < \Ra_2^*$, the Lyapunov exponent is nearly zero or negative, consistent with convergence to a fixed point. As $\Ra$ crosses $\Ra_2^*$, $\lambda$ suddenly grows large, indicating extreme sensitivity to initial conditions that is characteristic of chaotic dynamics. For $\Ra > 10^9$, $\lambda$ is nearly zero, reflecting a return to order.
}
\label{fig7}
\end{figure}

To further quantify the chaotic regime and distinguish it from the orderly dynamics, \cref{fig7} shows the fractal dimension $D_2$ (specifically the correlation dimension \cite{ott2002chaos}) and the Lyapunov exponent $\lambda$ corresponding to the trajectories shown in \cref{fig6}(a). The figure shows that at low $\Ra$, the fractal dimension is zero and the Lyapunov exponent is small or negative, consistent with convergence to a stable fixed point (either the conducting state or one of the circulating states). As $\Ra$ crosses $\Ra_2^* = 1.6 \times 10^7$, $D_2$ suddenly increases beyond unity and $\lambda$ grows large. For $\Ra$ in the range $[\Ra_2^*, 10^9]$, the value of $D_2$ indicates a strange attractor with dimension in between 1 and 2, while the large value of $\lambda$ indicates extreme sensitivity to initial conditions. Both observations suggest chaotic dynamics. As $\Ra$ crosses $10^9$, $D_2$ suddenly drops to approximately one and $\lambda$ drops to nearly zero, indicating a return to orderly dynamics, specifically periodic motion along a one-dimensional limit cycle.

\section{Return to order at high Rayleigh number}
\label{sec-pendulum}

\Cref{fig5,fig6,fig7} demonstrate that at very high Rayleigh number, large-scale order returns. The LSC reversals become periodic and the fluid CoM moves along an arc-like path, reminiscent of pendulum motion [see \cref{fig9}(a) for a close-up view]. In this section, we will reconcile this high-$\Ra$ behavior with the pendulum structure of \cref{Ldot*,xdot*,ydot*} that was observed in Section \ref{sec-lowdim}.

\begin{figure}[htb]
 \includegraphics[width=5.2in]{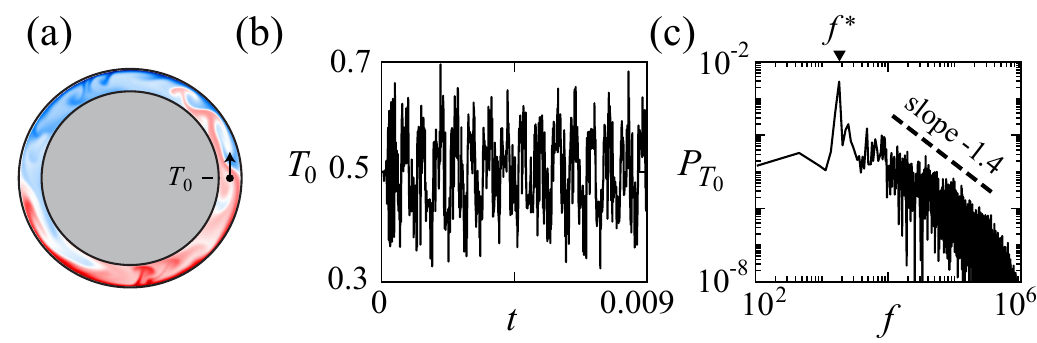}
  \caption{Temperature measurements reveal the presence of large-scale periodic motion and small-scale turbulent fluctuations. (a) A temperature ``probe" is placed at the center of the annular channel, $\theta=0$, $r=0.45$. (b) The temperature $T_0$ measured at the probe shows periodic oscillation in time, with small-scale perturbations due to the turbulent flow field. (c) The power spectrum of $T_0$ shows the main frequency of oscillation $f^*$ and a power-law decay with exponent $-1.4$ in the inertial regime of turbulence. Parameter values are $\Ra = 1.6\times10^9$, $\Pra = 4$, and $r_0 = 0.4$.}
\label{fig8}
\end{figure}

\begin{figure}[htb]
 \includegraphics[width=5.2in]{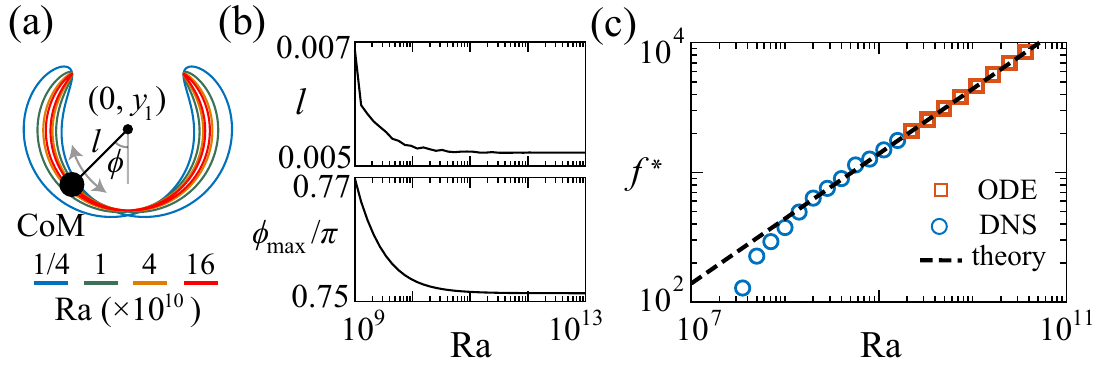}
  \caption{At high $\Ra{}$, the fluid CoM oscillates like a mechanical pendulum. (a) The CoM trajectory approaches a circular arc as $\Ra\to\infty$, showing periodic oscillation. (b) The pendulum length $l$ and maximum swing angle $\phi_{\max{}}$ each converge to an asymptote as $\Ra{} \to \infty$. 
  (c) The dominant frequency $f^*$ of LSC reversals measured from DNS and from the ODE model are well predicted by \cref{Period2} at high $\Ra$. 
  (a)-(b) and squares in (c) are obtained from the ODE solutions, circles in (c) are from full DNS. All data have $\Pra = 4$ and $r_0 = 0.4$.}
\label{fig9}
\end{figure}

First, although the large-scale dynamics are orderly at high Rayleigh number, \cref{fig8} shows that turbulent fluctuations prevail at the small scales. The snapshot seen in \cref{fig8}(a) illustrates the small-scale complexity of the temperature field arising in the DNS at $\Ra = 1.6 \times 10^9$. To further characterize this state, \cref{fig8}(b) shows a time-series of the temperature $T_0$ measured at a fixed location in the annulus.
The signal exhibits a dominant oscillatory structure, with period corresponding to the LSC reversals. However, the signal shows significant fluctuations about this periodic oscillation. The frequency power-spectrum of $T_0$ shown in \cref{fig8}(c) reveals greater detail. The spectrum peaks at a value $f^*$ corresponding to the main periodic component, and thus the frequency of LSC reversals. 
At higher frequencies, the spectrum decays with a -1.4 power, consistent with the Bolgiano-Obukhov turbulence scaling of natural convection \cite{Wu1990, Lohse2010}. These observations not only demonstrate the presence of turbulence at $\Ra = 1.6 \times 10^9$, but also confirm that the DNS successfully resolves the turbulent behavior.

We now ask whether the dominant frequency, $f^*$, of LSC reversals in the high-$\Ra$ regime can be predicted by the ODE model, \cref{Ldot*,xdot*,ydot*}? As observed in Section \ref{sec-lowdim}, this system corresponds to a damped, driven pendulum with fulcrum $(0,y_1)$, angle $\phi(t)$, mass $m$, length $l$, CoM coordinates $(\xc, \yc) = (l \sin \phi, \, y_1 - l \cos \phi)$, and angular momentum $L = m l^2 \dot{\phi}$. In the following analysis, we identify an energy law satisfied by this pendulum system that leads to accurate estimates of the frequency $f^*$.

First, the length of the pendulum can generally vary with time $l = l(t)$. Multiplying \cref{xdot*} by $\xc$, \cref{ydot*} by $\yc$, and adding gives the exact relation
\begin{equation}
\td{}{t}{l^2} = -2 \beta l^2 + 2\beta \dy (Y-y_1) ,
\end{equation}
where $\dy = y_0 - y_1 > 0$. For the system to reach a limit cycle, the time average of $dl^2/dt$ must vanish, giving the exact relationship
\begin{align}
\label{meanell2}
\mean{l^2} = \dy \mean{Y - y_1} ,
\end{align}
where $\mean{\cdot}$ indicates a time average. This equation offers an immediate observation: the fact that $l^2$ is non-negative implies that $\mean{\yc-y_1} \ge 0$. Hence, in the case of a limit-cycle solution, the average position of the CoM lies above the fulcrum.

Next, to estimate the oscillation frequency it is necessary to solve for $l$, at least in the mean sense. To this end, we introduce the energy
\begin{equation}
\label{energy}
E = \frac{1}{2} k L^2 + \Ra \Pra (\yc-y_1) .
\end{equation}
The terms on the right side of \cref{energy} represent kinetic and potential energy respectively. Taking a time derivative, using \cref{Ldot*,ydot*}, and simplifying yields the energy law,
\begin{equation}
\label{Edot}
\dot{E} = -\alpha \Pra \, k L^2 + \beta \Ra \Pra (y_0-\yc) .
\end{equation}
The first term on the right-hand side represents energy dissipation associated with a non-trivial flow field, $L \ne 0$ (recall $\Pra$ is proportional to viscosity). The second term represents positive energy injected into the system by the driving terms with prefactor $\beta$ in \cref{ydot*}.
For a limit cycle to exist, the condition $\mean{\dot{E}} = 0$ must hold, giving
\begin{equation}
\label{meanL}
k \alpha \mean{L^2} = \Ra\, \beta \mean{y_0 - \yc} .
\end{equation}
When this condition is met, the energy lost to dissipation balances the energy injected into the system over a cycle.

\Cref{meanell2,meanL} constitute two constraints for the three unknowns $\mean{l^2}, \mean{\yc}, \mean{L^2}$. One additional constraint is needed to close the system. To obtain this last constraint, we introduce two assumptions. First, we assme the length $l$ to be nearly constant in time, as is consistent with numerical measurements that will be shown in \cref{fig9}(a). Second, although \cref{Edot} shows that energy is not conserved in general, it is conserved on average for a limit cycle. We will therefore assume the energy to be equal to its average value $E(t) = \mean{E}$. Taking the time average of \cref{energy} gives 
\begin{equation}
\label{Eavg}
\mean{E} = \frac{1}{2} k \mean{L^2} + \Ra \Pra \mean{\yc-y_1} .
\end{equation}
At the bottom of the swing, $\yc = y_1-l$, the angular momentum is near it's maximum $L = L_{\max}$, giving energy
\begin{equation}
\label{Ebot}
\Ebot = \frac{1}{2} k \Lmax^2 - \Ra \Pra \, l .
\end{equation}
Assuming nearly constant energy, $\Ebot=\mean{E}$, gives the relationship
\begin{equation}
\label{Ematch}
\Ra \Pra \left( l + \mean{\yc-y_1} \right) = \frac{1}{2} k \left( \Lmax^2 - \mean{L^2} \right) .
\end{equation}
Naturally, the scales of $\Lmax^2$ and $\mean{L^2}$ are directly related. For example, if $L(t)$ varies sinusoidally, then $\mean{L^2} = \Lmax^2/2$. We therefore set $\mean{L^2} = \Lmax^2/m$ for some constant $m$ to be chosen later (e.g.~$m=2$ for a sinusoidal wave and $m=3$ for a triangular wave). Making this substitution in \cref{Ematch}, while using \cref{meanL}, and simplifying gives
\begin{equation}
2 \alpha \Pra \left( l + \mean{\yc-y_1} \right) = \beta (m-1) \mean{y_0 - \yc} .
\end{equation}
Then inserting \cref{meanell2} with constant $l$ gives a quadratic equation for $l$
\begin{equation}
2 \alpha \Pra \, l (\dy + l) = \beta (m-1) \left( \dy^2 - l^2 \right) .
\end{equation}
The quadratic can be factored exactly and possesses one positive root,
\begin{equation}
\label{leq}
l = \dy \left( \frac{(m-1) \beta}
{(m-1)\beta + 2 \alpha \Pra} \right) .
\end{equation}
We have therefore solved for the pendulum length under the assumptions that a limit cycle has been reached and that the length is nearly constant. The parameter $m$ relates the maximum and RMS values of angular momentum $L$. Observations from DNS suggest that $L(t)$ lies approximately between a sinusoidal $(m=2)$ and triangular $(m=3)$ waveform; see, for example, the right-most panel of \cref{fig2}. We will therefore set $m=2.5$.

If the amplitude of the pendulum motion were small, we could determine the period $T_p$ right away using the well-known formula $T_p = 2\pi \sqrt{l/g}$, where $g = k l^2 \, \Ra \, \Pra$ for the pendulum system given by \cref{Ldot*,xdot*,ydot*}. Numerical measurements, however, will show the amplitude of motion to be large [see \cref{fig9}(a)]. In this case, the period is given by the more general formula 
\begin{equation}
\label{Period}
T_p = \frac{4}{\sqrt{k l \,\Ra\,\Pra\,}} \, K \left( \sin^2\frac{\phim}{2} \right),
\end{equation}
where $\phim$ is the maximum angle reached by the pendulum, and $K(x) = \int_0^{\pi/2} \left(1-x^2\sin^2\theta\right)^{-1/2} d\theta$ is the complete elliptic integral of the first kind. The value of $\phim$ is thus needed to estimate the period and hence $f^*$.

At the apex, $\phi = \phim$ and $\dot{\xc}=\dot{\yc}=0$, which upon inserting into \cref{xdot*,ydot*} and simplifying gives $ \xc^2 = (y_0-\yc)(\yc-y_1)$. Inserting the definitions $\xc = l \sin\phi$ and $\yc = y_1 - l \cos \phi$ and solving for $l$ gives the relationship $l = -\dy \cos \phim$. Some further manipulations then give the argument of the elliptic integral in \cref{Period} as
\begin{equation}
\sin^2 \frac{\phim}{2} = \frac{1-\cos \phim}{2} = \frac{\dy + l}{2 \dy} = \frac{(m-1)\beta + \alpha \Pra}{(m-1)\beta + 2\alpha \Pra} ,
\end{equation}
where we have assumed constant $l$ as given by \cref{leq}. Inserting this formula into \cref{Period} gives the period of oscillations,
\begin{equation}
\label{Period2}
T_p = \frac{4}{\sqrt{k l \,\Ra\,\Pra\,}} \, K \left(  \frac{(m-1)\beta + \alpha \Pra}{(m-1)\beta + 2\alpha \Pra}\right) .
\end{equation}
The frequency of LSC reversals is then given by $f^* = 1/T_p$. 

\Cref{fig9} shows numerical measurements of the CoM motion in the high-$\Ra$ regime in comparison to this prediction. First, \cref{fig9}(a) shows CoM trajectories computed numerically from \cref{Ldot*,xdot*,ydot*} for Rayleigh numbers in the range $\Ra = 1/4 \mbox{--} 16 \times 10^{10}$. Each trajectory closely resembles the rhythmic swinging of a pendulum about the fulcrum point $(0,y_1)$ that is predicted by \cref{y1}. At $\Ra = 1/4 \times 10^{10}$ (blue orbit), the pendulum length $l$ varies somewhat over the period. At higher Rayleigh number, though, the orbit tightens and $l$ remains nearly constant throughout the period. This observation is consistent with the assumption of constant $l$ made in the analysis above. \Cref{fig9}(b) show measurements of the pendulum length $l$ and maximum swing angle $\phim$, both of which lie in a relatively narrow range over four decades of $\Ra$. Interestingly, as $\Ra \to \infty$, $\phim$ appears to converge to a value near $3\pi/4$.

Most importantly, \cref{fig9}(c) shows numerical measurements of the LSC reversal frequency $f^*$ in comparison to the theoretical prediction \cref{Period2}. The figure shows measurements of $f^*$ taken from both the DNS (blue circles) and from simulation of the ODE system (orange squares), along with the prediction from \cref{Period2} with $m=2.5$ (dashed line). The figure shows that \cref{Period2} accurately predicts the LSC reversal frequency over roughly the largest decade of Rayleigh numbers that are practical for DNS. For $\Ra > 2 \times 10^9$, the DNS becomes computationally prohibitive, but measurements of $f^*$ from simulations of the ODE model are possible and still agree with the prediction from \cref{Period2}. The close agreement between DNS, the ODE model, and \cref{Period2} suggest that the main mechanism for high-$\Ra$ LSC reversals has been properly accounted for. In particular, LSC reversals result from an inertial overshoot of the CoM, directly analogous to a damped, driven pendulum system.

\section{Discussion}
\label{sec-discussion}

In this work, we have examined thermal convection in an annulus using both DNS and a simplified ODE model that derives systematically from the governing equations. In both the DNS and the ODE model, we observe the onset of fluid motion at a critical Rayleigh number $\Ra_1^*$, where flow begins to circulate steadily in one direction, and we observe the transition to chaotic bidirectional flows at a higher critical Rayleigh number $\Ra_2^*$. Stability analysis of the ODE model yields formulas for $\Ra_1^*$ and $\Ra_2^*$ that accurately predict these transitions with no adjustable parameters, demonstrating a modeling accuracy that has not been achieved previously. Both the DNS and ODE model show a high-Rayleigh number state, in which the bulk flow changes direction periodically despite small-scale turbulent fluctuations in the flow field.

There are still many interesting aspects of this annular convection problem awaiting exploration. First, one surprising observation in \cref{fig6} is the existence of a critical $\Pr^*$, below which the circulating state remains stable for arbitrarily large $\Ra$. For the case shown in \cref{fig6}, the threshold value $\Pr^*=0.55$ is below that of common working fluids such as water. However, liquid metal convection \cite{ren2022flow} is known to have low $\Pra{}$ due to the high thermal conductivity. Thus, future experiments featuring liquid metal could attempt to verify this ever-circulating state. Such experiments, combined with new analysis that builds upon the present theory, could offer insight into thermal transport and potentially novel $\Nu$--$\Ra$ scaling relationships associated with the ever-circulating state.

Secondly, in this work, we have examined a fixed annular geometry with $r_0 = 0.4$, but changing the radius of the inner boundary could certainly affect the states of convection.  Preliminary inquiries suggest smaller $r_0$ causes the periodic state to appear over a wider range of $\Ra$. In the limit of $r_0 \to 0$, the annular geometry tends to a circular domain, which is a canonical case worthy of study. We note that the low-dimensional ODE model discussed here loses accuracy in this limit as a result of the wider channel permitting larger deviations from shear flow. This situation thus presents new modeling challenges for future work.

\section*{\si{}} Supplementary movies are available at \url{https://math.nyu.edu/~jinzi/research/AnnularConvection/Movie/}.
\bibliography{manuscript}

\end{document}